\documentclass[prd, twocolumn, nofootinbib]{revtex4}
\usepackage{epsfig}

\newcommand{\beq}{\begin{equation}}
\newcommand{\eeq}{\end{equation}}
\newcommand{\beqa}{\begin{eqnarray}}
\newcommand{\eeqa}{\end{eqnarray}}
\newcommand{\om}{\Omega_m}

\begin{document} 

\title{Strong Gravitational Lensing and Dark Energy Complementarity} 
\author{Eric V.~Linder} 
\affiliation{Physics Division, Lawrence Berkeley National Laboratory, 
Berkeley, CA 94720} 

\begin{abstract} 
In the search for the nature of dark energy most cosmological probes 
measure simple functions of the expansion rate.  While powerful, these 
all involve roughly the same dependence on the dark energy equation of 
state parameters, with anticorrelation between its present value $w_0$ 
and time variation $w_a$.  Quantities that have instead positive 
correlation and so a sensitivity direction largely orthogonal to, e.g.,  
distance probes offer the hope of achieving tight constraints 
through complementarity.  Such quantities are found in strong 
gravitational lensing observations of image separations and time delays. 
While degeneracy between cosmological parameters prevents full 
complementarity, strong lensing measurements to 1\% accuracy can 
improve equation of state characterization by 15-50\%.  Next generation 
surveys should provide data on roughly $10^5$ lens systems, though 
systematic errors will remain challenging. 
\end{abstract} 

\maketitle 

\section{Introduction} \label{sec.intro}

Dark energy poses a fundamental challenge to our understanding of the 
universe.  The acceleration of the cosmic expansion discovered through 
the Type Ia supernovae distance-redshift relation can be interpreted 
in terms of a new component of the energy density possessing a 
substantially negative pressure.  Further observations from the cosmic 
microwave background radiation indicate the spatial geometry of 
the universe is flat; in combination with the supernova measurements 
this implies that the unknown ``dark energy'' comprises roughly 70\% of 
the total density, in concordance with large scale structure data 
indicating that matter contributes approximately 30\% of the critical density. 

Such a weight of dark energy leads to its dominance of the expansion, causing 
acceleration, restricting the growth of large scale structure, and holding the 
key to the fate of the universe.  But apart from its rough magnitude, its 
nature is almost unknown -- whether it arises from the physics of the 
high energy vacuum, a scalar field, extra dimensional or ``beyond Einstein'' 
gravitational effects, etc.  One way to gain clues to the underlying 
physical mechanism is to characterize the behavior of dark energy in terms 
of its equation of state ratio as a function of 
redshift (essentially a time parameter), $w(z)$.  This is conventionally 
interpreted in terms of the ratio of the dark energy pressure to energy 
density, but can also be used as an effective parameterization to treat 
generalizations of the cosmology framework of the Friedmann equations 
of expansion \cite{linjen} or 
simply in terms of the expansion rate and acceleration itself \cite{lingrav}. 

To obtain the clearest focus on the class of physics responsible for the 
accelerating universe, we seek the tightest constraints on the equation 
of state.  This must come not merely from high statistical precision of 
a probe, but from robust control of systematic uncertainties.  The 
greatest accuracy and confidence 
in the measurements will come from independent crosschecks and complementarity 
between methods of probing the cosmology.  Many studies have considered such 
complementarity between probes (e.g.\ \cite{fhlt,linap,seo,linbo,wl,cluster}), 
with a 
promising future for next generation surveys carrying out such measurements. 
However, all the methods of supernova distances, cosmic microwave background 
power spectrum, weak gravitational lensing, cluster counts, and 
baryon oscillations 
possess a similar fundamental dependence on the equation of state through the 
Hubble parameter, or expansion rate. 

This paper investigates whether a truly complementary probe exists that has 
nearly orthogonal dependence to the previous ones in the plane of the 
equation of state parameters of value today, $w_0$, vs.\ time variation, 
$w_a$.  Such a probe, if practical, would offer a valuable contribution 
to uncovering fundamental physics 
and deserve further consideration among next generation experiments.  Section 
\ref{sec.anti} considers the characteristics such a method would possess and 
identifies two promising candidates related to strong gravitational 
lensing.  In \S\ref{sec.fisher} we analyze the 
sensitivity of these probes to the cosmological parameters and the constraints 
and complementarity they offer.  In the conclusion we summarize the 
prospects for strong lensing as a cosmological probe and discuss some issues 
regarding surveys and systematic uncertainties. 

\section{Complementarity in the Equation of State Plane} \label{sec.anti} 

Astronomical observations involving distances and volumes all follow from the 
metric in a simple, kinematic way \cite{weinberg}.  For convenience, 
the dependence 
can be written in terms of the conformal distance interval 
\beq 
d\eta=dt/a(t)=-dz/H(z), 
\eeq 
where $a(t)=(1+z)^{-1}$ is the scale factor as a function of proper time, 
equivalently parametrized in terms of redshift $z$, and $H=\dot a/a$ is 
the Hubble parameter.  So in observing some source at redshift $z$ the 
cosmological information is carried by the quantity $H$. 

Varieties of distance observations -- angular diameter, luminosity, 
proper motion -- all contain the same information.  One exception is 
the parallax distance (cf.\ \cite{fpoc}, \S3.2) which involves not only 
$H$ but the spatial curvature $k$ as well; however this is not practical 
for cosmology and CMB measurements strongly indicate a flat universe.  
Another subtlety involves exotic models that break the thermodynamic, or 
reciprocity, relation between the various distances 
\cite{thermo,tolman,bassett} by violation of Liouville's theorem, but 
this arises from photon properties and not cosmology. 

Volume elements, and hence numbers of sources, are built up out of 
distances and so similarly involve the Hubble parameter.  From the 
Friedmann equation the relation between the expansion rate and the 
dark energy equation of state is 
\beq 
H^2/H_0^2=\om(1+z)^3+\Omega_{DE} e^{3\int d\ln(1+z)\,[1+w(z)]}, \label{eq.hw}
\eeq 
where $H_0$ is the present value of the Hubble parameter (the Hubble 
constant), and for a flat universe $\Omega_{DE}=1-\om$.  As discussed 
in \cite{linjen} this can be written more generally as defining 
an effective equation of state 
\beq 
w(z)=-1+\frac{1}{3}\frac{d\ln\delta H^2/H_0^2}{d\ln(1+z)}, 
\eeq 
where $\delta H^2$ encodes our ignorance of the right hand side of 
eq.\ \ref{eq.hw} after the first, matter density term. 

Making $w(z)$ more positive (holding $H_0$ and $\om$ fixed) increases 
the expansion rate in the past, and so decreases the acceleration (since 
the rate today is fixed).  Distances will be less, sources appear 
brighter, etc.  If we try to characterize the nature of the dark energy 
in the simplest, nontrivial way, through the value of its equation of 
state today and its increasing or decreasing positivity into the past, 
we see that cosmological measurements, depending only on $H$, don't 
really care in a gross sense {\it how} $w(z)$ was more positive -- 
either increasing the present value $w_0$ or increasing the time 
variation $w_a\sim dw/d\ln(1+z)$ does the trick. 
Thus, for all such probes the quantities $w_0$, $w_a$ will be 
anticorrelated; an increase in one can be (at least partially) offset 
by a decrease in the other.  If we plot constraints from astronomical 
data in the $w_0-w_a$ plane (marginalizing over any other parameters), 
the probability contours will show a degeneracy direction 
tilted counterclockwise from the vertical.  This is well known and 
illustrated for a number of different probes in \cite{hutcoo}; note 
also that a ``bare'' $H$ of course has the same dependence \cite{linbo}. 

However, this has the fundamental implication that complementarity 
between methods can only be partial in this physically key parameter 
plane; different data sets cannot be orthogonal.  This restricts the 
ability to constrain the equation of state and understand the nature 
of dark energy.  What would be valuable is construction of a cosmological 
probe whose sensitivity lies clockwise of vertical, into the unexplored 
half-plane of positive correlation between $w_0$ and $w_a$.  Note that 
this argument does not depend on the specifics of how $w_0$, $w_a$ are 
defined, but for concreteness we adopt the parametrization 
\beq 
w(z)=w_0+w_a(1-a)=w_0+w_a z/(1+z), 
\eeq 
successful in fitting smoothly varying dark energy models 
\cite{linprl,lin10217}.  A characteristic time variation of the equation 
of state is given by $w'\equiv dw/d\ln(1+z)|_{z=1}=w_a/2$. 

\begin{figure}[!hbt]
\begin{center} 
\psfig{file=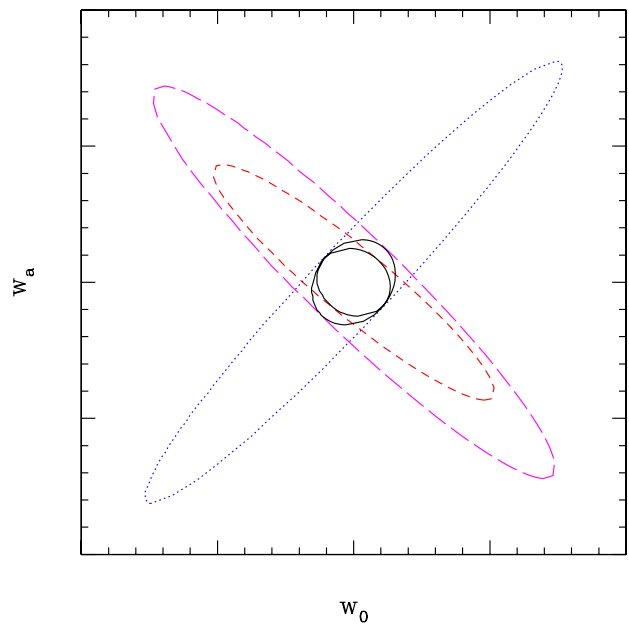,width=3.4in} 
\caption{This cartoon illustrates the value of orthogonality in the 
equation of state plane for tightly constraining dark energy 
parameters.  The short dashed ellipse represents a distance-redshift 
probe, or any other that has an anticorrelation between the equation 
of state parameters.  Adding a positive correlation probe (dotted) 
gives a joint contour (inner solid curve) with considerably smaller 
uncertainties.  
This also helps to reduce the impact of systematics (long dashed 
curve and outer solid curve).  However the situation is more complicated 
when marginalizing over a larger set of parameters. 
} 
\label{fig.compl}
\end{center} 
\end{figure}

Figure \ref{fig.compl} illustrates this anticorrelation for the 
distance-redshift probe (dashed ellipse).  A hypothetical probe that 
involves positive correlation between the equation of state parameters 
(dotted ellipse) can give nearly orthogonal constraints and so the 
joint probability contour (solid ellipse) between the two types of 
probe tightly characterizes the equation of state.  Furthermore, since 
reduced data quality, e.g.\ from systematic uncertainties, often 
lengthen the confidence contour along the degeneracy direction (long 
dashed ellipse), an orthogonal probe immunizes against systematic 
errors.  However, degeneracies with other cosmological parameters 
interfere with both these desiderata, as discussed in \S\ref{sec.fisher}. 

The growth of large scale structure, or equivalently the evolution 
of gravitational potentials, and hence, say, the number of galaxy 
clusters of a certain mass, might seem to offer a different dependence 
than distances.  But the growth equation also has $H$ as its predominant 
ingredient and indeed in \cite{hutcoo} we see that the growth factor 
still possesses the anticorrelation.  An exception might occur for 
models where the dark energy (or alternative gravity theory) is 
inhomogeneous itself and can act as a source to the matter density 
perturbations, but this is not expected to occur for scalar field 
dark energy on subhorizon scales \cite{caldwell,dave}. 

If simple measures of distances or products of distances all give 
anticorrelation, yet they involve differing levels of sensitivity.  
This provides the hope that opposing these quantities through ratios 
might break the pattern of degeneracy.  Indeed this was found in 
\cite{linap} on consideration of the cosmic shear test\footnote{A note 
on the name: this probe involves the observed shearing of 
a sphere due to the properties of spacetime itself.  Despite a perfectly 
isotropic space, a sphere will appear sheared because the spacetime is not 
isotropic -- directions along and perpendicular to the null geodesics differ. 
This shear is due to cosmic properties, unlike shears in gravitational lensing 
due to anisotropies not in spacetime but in space, though the latter is 
unfortunately sometimes called cosmic shear.} 
originated by \cite{alpac}.  Here the key quantity was 
$H(z)d_A(z)\sim H(z)\int dz/ H(z)$, where $d_A$ is the angular diameter 
distance.  For a certain redshift range, $z\approx1.3-2.3$ (for $\om=0.3$), 
the competition between the two ingredients causes a positive 
correlation between $w_0$ and $w_a$.  A critical pitfall however was 
that systematic uncertainties could blow up the constraint contour in 
a preferred direction such that the resulting contour actually exhibited 
an anticorrelation (compare Figs.\ 5 and 8 of \cite{linap}).  Another 
indication of positive correlation was found between discrete values 
of $w(z)$ in neighboring redshift bins by \cite{songknox} for 
ratios of distances entering weak gravitational lensing calculations. 

Here we investigate distance ratios that appear in strong gravitational 
lensing measurements.  Specifically, we consider $r_{ls}/r_s$, related 
to the angular separation $\Delta\theta$ between multiple images of a 
source, $r_{ls}/(r_lr_s)$, related to the time delay $\Delta t$ between 
the images, and $r_lr_{ls}/r_s$, connected with the length scale of 
caustic properties, or the cross section, of the lensing.  Here $r_l$ 
is the comoving distance to the lens, $r_s$ to the source, and 
$r_{ls}$ between the source and lens; in a flat universe $r_{ls}=r_s-r_l$. 

\begin{figure}[!hbt]
\begin{center} 
\psfig{file=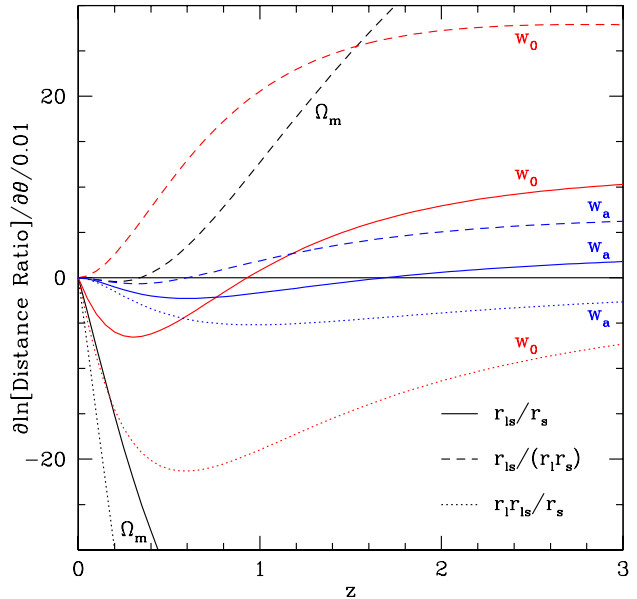,width=3.4in} 
\caption{The sensitivity of the distance ratios to the cosmological 
parameters $\theta=\{\om,w_0,w_a\}$ are encoded in the derivatives 
plotted here, normalized to 1\% fractional measurement of the distance 
ratio.  The larger the absolute 
magnitude of the derivative at a particular redshift, the more 
constraining the observations there. 
} 
\label{fig.distratio}
\end{center} 
\end{figure}

Figure \ref{fig.distratio} illustrates the sensitivity of each of these 
ratios to the cosmological parameters as a function of redshift.  
We have taken the redshift to represent the lens redshift $z_l$ and 
for convenience fixed the source redshift to be $z_s=2z_l$ (since that 
gives roughly the greatest probability of effective lensing).  Generally 
the greatest 
sensitivity is to the magnitude of the matter (or dark energy) density, then 
to the present value of the dark energy equation of state ratio, and least to 
the time variation of the equation of state.  

Of most interest to our 
investigation are the crossings from negative to positive sensitivity, 
i.e.\ at some redshift the distance ratio switches from diminishing 
to growing as we change the value of a parameter $\theta=\{\om,w_0,w_a\}$ 
(increasing it, say).  
Since these crossings occur at different redshifts if the parameter is 
$w_0$ than if it is $w_a$, then the correlation between these 
quantities can shift from negative to positive.  We see that positive 
correlation occurs for $r_{ls}/r_s$ for $z_l\approx 0.9-1.7$ and for 
$r_{ls}/(r_lr_s)$ 
for $z_l\approx 0-0.6$; zero crossing in equation of state variables 
and hence positive correlation 
does not occur for $r_lr_{ls}/r_s$.  (The exact values will depend on 
the fiducial cosmology, here taken to be $\om=0.3$, $w_0=-1$, $w_a=0$.) 

\begin{figure}[!hbt]
\begin{center} 
\psfig{file=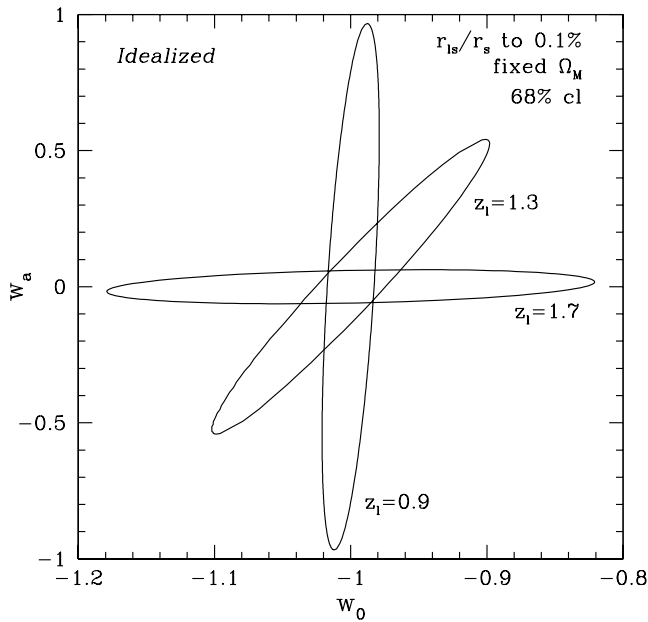,width=3.4in} 
\caption{Confidence contours for an idealized experiment measuring 
the ratio $r_{ls}/r_s$ rotate depending on the redshift.  Degeneracy 
directions follow the Fisher sensitivities from Fig.\ \ref{fig.distratio}, 
with zero crossings of a parameter giving contours parallel to that axis. 
Note the strong positive correlation between equation of state variables 
for $z_l=1.3$, and hence strong complementarity with, e.g., distance 
measurements. 
} 
\label{fig.slflower}
\end{center} 
\end{figure}

For the ratio $r_{ls}/r_s$ we expect rapid evolution in the $w_0-w_a$ 
degeneracy direction as the contour will rotate from horizontal at 
$z=0.9$ (no sensitivity to $w_0$, so lying parallel to the $w_0$-axis) 
to vertical at $z=1.7$ (no sensitivity to $w_a$).  Fig.\ \ref{fig.slflower} 
shows this behavior in a ``flower'' plot, with idealized precision 
to better illustrate the degeneracy direction.  A strong lensing survey 
measuring image separations from lenses within this redshift range 
should possess good internal complementarity and the ability to constrain 
the dark energy equation of state.  Furthermore, in this 
range strong lensing will have near orthogonality with supernova distance and 
weak gravitational lensing surveys and so be a valuable complementary probe.  

The ratio $r_{ls}/(r_lr_s)$ appears somewhat less promising.  Its 
sensitivity to $w_a$ is much weaker in the key redshift range and 
the evolution of the correlation between $w_0$ and $w_a$ much less 
dramatic.  In particular the contour never tilts far from the vertical.  
This means it is not as complementary either internally or with other 
cosmological probes.  But it does possess three interesting 
characteristics: its overall sensitivity to the equation of state is 
reasonably large at $z>1$, the positive correlation region is at low 
redshifts where observations are easier (though there is 
less volume, hence fewer lens systems), and it has an odd null 
sensitivity with respect to $\om$ at $z=0.35$.  This last is where its 
sensitivity to $w_a$ is maximal in the positive correlation region and 
eases degeneracy with the matter density and hence reduces the need for 
a tight external prior on it.  So time delay measurements should 
not be wholly neglected as a possibly useful probe. 

\section{Constraints on Cosmological Parameters} \label{sec.fisher} 

So far we have only concentrated on the degeneracy direction.  What about 
the actual magnitude of constraints capable of being imposed on the 
cosmological parameters?  If we consider one parameter $\theta$ at a time, 
fixing the others, we can obtain a lower limit 
on the parameter estimation uncertainty.  This is given by 
\beq 
\delta\theta\ge \left|\frac{\partial[{\rm Distance\ Ratio}]}{\partial\theta} 
\right|^{-1} \delta[{\rm Distance\ Ratio}]. 
\eeq 
In Fig.\ \ref{fig.distratio} we took a fractional measurement error of 1\%.  
Thus, for example, the sensitivity for $r_{ls}/r_s$ with respect to the 
parameter $w_0$ at $z=0.3$ is -6.5 so the best possible estimate of $w_0$ 
from such a single, 1\% measurement is 1/6.5=0.15.  That is, the 
unmarginalized uncertainty is $\sigma(w_0)= 0.15$.  This will be degraded 
by degeneracies with other parameters, or by worse precision in the 
measurement, and improved by additional measurements, subject to some 
systematics floor.  

In fact, degeneracies with additional parameters wash out the simple 
orthogonality of Fig.\ \ref{fig.compl}.  The parameter phase space is 
actually 3-dimensional (from $\om$) or higher and the joint probabilities 
do not merely trace the intersection of the contours in the equation of 
state plane.  This weakening of apparent complementarity also means 
that the magnitude of the sensitivity with redshift matters as well 
as the degree of correlation, and this can alter the favored redshift 
range.  All these effects need to be taken into account. 

Here we seek to establish what improvements the complementarity of strong 
lensing data with other probes such as distance-redshift or weak lensing 
data can realistically provide.  Next generation precision in all 
three of these come in one 
package: the proposed Supernova/Acceleration Probe (SNAP: \cite{snap}).  With 
some $10^7$ galaxies observed in its deep survey (down to AB mag $R=30$) 
and $10^8$ in its wide survey (to AB mag $R=27.5$), the canonical 
estimate of one strong lens system per thousand galaxies provides an 
anticipated wealth of data.  Other valuable data sets will come from 
LSST \cite{lsst} and LOFAR \cite{lofar}. 

\subsection{Image Separation} \label{sec.sep} 

If we consider an optimistic strong lensing measurement precision of 1\% 
in a redshift bin of 0.1, whether 
through statistical means, e.g.\ 100 lens systems of 10\% precision, or as a 
systematic floor, we find that the parameter uncertainties are large if we 
do not a priori fix some of the variables.  Even employing the orthogonal 
combination of measurements of the strong lensing separation variable 
$S=r_{ls}/r_s$ at $z=0.9$ and $z=1.7$ (cf.\ Fig.\ \ref{fig.slflower}), the 
estimations are $\sigma(w_0)=0.15$, $\sigma(w_a)=0.53$ even with $\Omega_m$ 
fixed, and 0.28, 0.93 with a prior on $\om$ of 0.03.  The lack of sensitivity 
of the strong lensing separation to the dark energy equation of state 
and, as mentioned above, the degeneracy with the other parameter, $\om$, 
prevent as 
well its complementarity with other probes from being as effective 
as might have been hoped from \S\ref{sec.anti}. 

\begin{figure}[!hbt]
\begin{center} 
\psfig{file=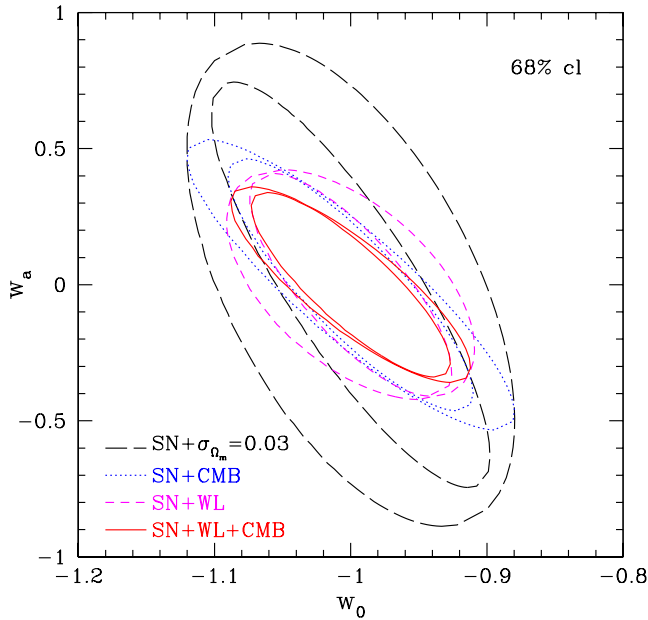,width=3.4in} 
\caption{Estimates of the dark energy equation of state parameters 
from various combinations of cosmological probes (SN=supernovae, 
WL=weak lensing, CMB=cosmic microwave background) improve modestly 
with the addition of strong lensing image separation measurements. 
Outer contours of each pair of line types include the labeled data 
sets as will be provided by SNAP and Planck; inner contours add the 
1\% strong lensing constraint. 
} 
\label{fig.slangle}
\end{center} 
\end{figure}

Figure \ref{fig.slangle} shows contours in the $w_0-w_a$ plane, now 
marginalized over $\om$, for various combinations of probes.  This considers 
1\% measurements at $z=0.9$, 1.3, 1.7.  The improvements in parameter 
estimation remain modest: about 15\% in $w_0$ and $w_a$ when added to 
SNAP supernova distance data, 19\% and 3\% respectively when added to the 
supernova plus weak lensing data set, 25\% and 13\% when added to 
supernova plus Planck CMB data, and 17\% and 6\% when added to 
all three data sets.  If strong lensing $S$ measurements were available from 
$z=0.1-1.7$, say (recall this is the lens redshift, with the source assumed to 
lie at twice this redshift), the improvement to the triple data set 
increases to 30\% and 15\%. 
If the measurement floor for the strong lensing was instead at 2\%, then the 
constraints in conjunction with the other three probes only gain due to strong 
lensing by 6\% and 2\% for $z=0.9$, 1.3, 1.7 or 17\% and 7\% for $z=0.1-1.7$. 

\subsection{Time Delay} \label{sec.time} 

The other strong lensing quantity we consider is $T=r_{ls}/(r_lr_s)$, related 
to the time delay between images.  This introduces an extra parameter -- 
the Hubble constant $H_0$, which we marginalize over.  Again, measurements 
of strong 
lensing by itself provide only poor estimates of the cosmological parameters, 
even for combinations of redshifts where the degeneracy directions are 
complementary.  The insensitivity to $\om$ at low redshift noted in 
\S\ref{sec.anti} does not help since the overall weak dependence and 
degeneracy of the equation of 
state variables prevents precision constraints. 

In complementarity with supernova distances, strong lensing $T$ measurements 
of 1\% precision at $z=0.1-0.6$ provide improvement by 21\% and 9\% in 
$w_0$ and $w_a$.  Adding them to all three other data sets helps by 
13\% and 9\%, or 22\% and 15\% if the strong lensing measurements 
extend from $z=0.1-1.7$.  Weakening the precision to 2\% adjusts this 
last case to 7\% and 5\%.  Note that no external prior on 
$H_0$ is required; it is determined from the data to better than 1\%. 

\subsection{Image Separation and Time Delay} \label{sec.septime} 

\begin{figure}[!hbt]
\begin{center} 
\psfig{file=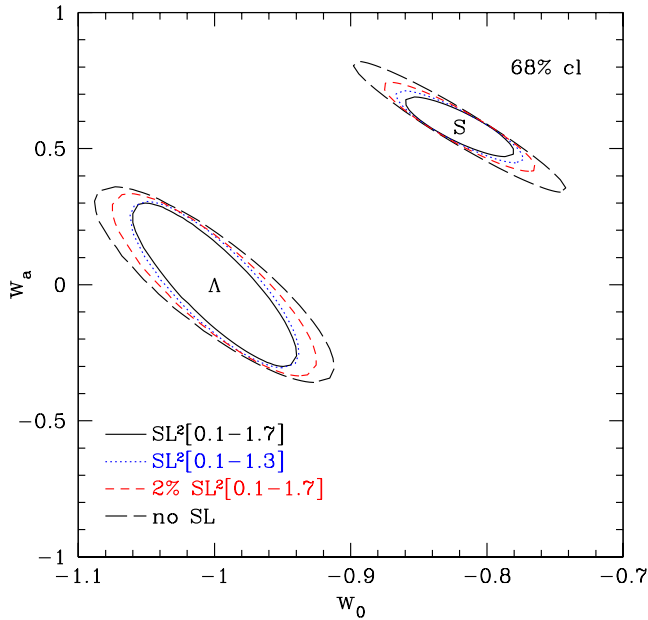,width=3.4in} 
\caption{Estimates of the dark energy equation of state parameters from 
supernovae, weak lensing, and CMB measurements can be further tightened 
by adding observations of both strong lensing quantities (SL$^2$): image 
separations and time delays.  The outer, long dashed ellipse with no strong 
lensing and ``$\Lambda$'' in its center corresponds to the outer solid 
ellipse of Fig.\ \ref{fig.slangle}. 
Strong lensing data at 1\% precision adds valuable complementarity (solid 
ellipse), even over a reduced redshift range (dotted).  Higher 
systematic uncertainties on strong lensing (2\%: short dashed), however, 
strongly diminish its usefulness.  As usual (cf.\ \cite{linbo}), dark 
energy models with time varying equation of state such as the SUGRA 
model (ellipses centered around ``S'') show enhanced complementarity. 
} 
\label{fig.snsl2}
\end{center} 
\end{figure}

Simultaneous addition of both separation and time delay information allows 
further improvements.  Relative to the supernova only data, these are 36\% and 
41\% for 1\% strong lensing data at $z=0.1, 0.3, \dots 1.7$; added to all 
three other probes they yield tighter constraints by 31\% and 17\%.  With only 
2\% precision the improvements in the latter case are 15\% and 7\%.  We 
show the parameter constraints in Fig.\ \ref{fig.snsl2} for the three  
probes without strong lensing, with strong lensing for the fiducial 
redshift bins, for a reduced strong lensing redshift range, and for the 
weaker 2\% strong lensing precision.

The highest redshift bins are not that powerful but it is important to include 
the $z=0.1$ bin to constrain $H_0$, which is strongly degenerate with 
$w_0$ and $w_a$.  Even with $H_0$ known to better than 1\% from the data, 
the degeneracy still plays an important role since the strong lensing 
time delay probe is so much more sensitive to $H_0$ than to the equation of 
state. 

As usual, taking the fiducial cosmology to be that of a cosmological 
constant tends to underestimate both the sensitivity and complementarity 
of the probes \cite{linbo}.  For comparison Fig.\ \ref{fig.snsl2} 
includes contours for a supergravity inspired dark energy model with 
time varying equation of state: S(UGRA) with $w_0=-0.82$, $w_a=0.58$. 
Here strong lensing offers more dramatic improvements, by 49\% in 
estimating $w_0$ and 54\% in $w_a$ 
for 1\% precision and 29\% and 22\% for 2\% precision.  Even at 2\% 
precision the final uncertainty from the four probes in complementarity 
achieves an impressive constraint on the time variation of 
$\sigma(w')=0.05$. 

\section{Conclusion} \label{sec.concl} 

While strong lensing does not completely fulfill the promise of orthogonal 
constraints on the dark energy equation of state that would provide vastly 
improved parameter estimates, it does offer some complementarity, 
especially for models with time varying equation of state.  
Furthermore, the data resources required will be innate within the deep 
and wide optical and near infrared surveys of SNAP; abundant strong lensing 
data should also come from LOFAR  at radio wavelengths and both strong and 
weak lensing from LSST in the optical.  We find that strong lensing 
adds reasonable value 
to dark energy constraints when the measurements reach the 1\% precision 
level.  This is in accord with other calculations (e.g.\ 
\cite{yamamoto,lewisibata}).  Precision arises from statistics 
subject to a systematics floor. 

The challenge in the next generation will not be the search for sufficient 
strong lensing data, it will be achieving systematics limits sufficiently 
low to allow the statistical wealth and the value of strong lensing 
complementarity to come into play.  Currently 1\% sounds rather optimistic 
for a bound to systematic uncertainties.  The main contribution is 
likely to come from our ignorance of the lensing mass model.  For example, 
within a singular isothermal sphere model 1\% measurement of the 
distance ratio $S$ requires knowledge of the velocity dispersion to 
0.5\% \cite{dhk}.  Cross correlation of different sources with the same 
lens to remove systematics, {\` a} la \cite{bernjain} for weak lensing, 
would not work, both because of the rarity of strong lensing and because 
sources at different distances and positions would probe different parts 
of the lensing mass distribution. 

One might hope to constrain the lensing mass distribution by making 
use of the simultaneous weak lensing 
information that SNAP or LSST would provide, but this will be of little use 
according to \cite{dalal}.  Basically the weak lensing measurements of 
shears average over a larger range of scales (or multipoles) than the 
convergence contribution from the mass to the strong lensing.  It is 
like trying to locate a pin with thick gloves.  Nongaussian effects 
also need to be incorporated in a more rigorous treatment of strong 
lensing.  Relaxation 
of the simplification that sources lie at redshifts $z_s=2z_l$ is 
unlikely to change significantly the results shown here. 

The advantage of image separations and time delays is that they are 
well defined and used as markers of the cosmic 
geometry.  One could also use the statistics of lensing systems, 
e.g.\ the number of lenses out to some redshift (e.g.\ 
\cite{sarbu,oguri,kuhlen}), 
but then one must deal with astrophysical scatter and observational 
bias and incompleteness.  Furthermore, 
lensing counts depend on mass density growth and volume factors 
that take us further from our original goal of finding orthogonal 
probes in the equation of state plane. 

Since systematics are the limiting factor, we can attempt to seek out 
special lensing systems that ameliorate these uncertainties, 
such as strongly lensed calibrated candles 
(e.g.\ Type Ia supernovae \cite{holz,ogurik}), time evolving, differentially 
amplified sources (e.g.\ Type II supernovae \cite{goobar,wagoner}), or 
relatively clean Einstein cross or ring images.  With the wealth of 
future data such subsampling may be practical. 

While interrelations between cosmological parameters entering into 
distances and the expansion rate prevent clean orthogonality between 
cosmological probes, the various methods retain sufficient complementarity 
to both crosscheck each other and further tighten constraints on the 
nature of dark energy.  If the systematics challenge can be met, in 
special systems at least, strong lensing merits further consideration 
in joining the cosmological toolbox of next generation probes.

\section*{Acknowledgments} 

I thank Dragan Huterer for helpful comments.  This work has been supported 
in part by the Director, Office of Science, Department of Energy under 
grant DE-AC03-76SF00098.

\end{document}